\documentclass[
twocolumn,
showpacs,superscriptaddress,a4paper, aps, longbibliography]{revtex4-1} %% use 11pt for Applied Optics

\usepackage[USenglish]{babel}

\usepackage{graphicx,xcolor}
\usepackage{amsmath, bbm}
\usepackage{natbib}
\usepackage{hyperref}
\usepackage{epstopdf}
\usepackage{color}

\renewcommand{\Im}{\mbox{Im}}
\renewcommand{\Re}{\mbox{Re}}

\renewcommand{\imath}[0]{\mathrm{i}}

\newcommand{\unit}[1]{\mathrm{#1}}
\renewcommand{\vec}[1]{\boldsymbol{#1}}

\newcommand{\myref}[1]{}%{{\color{blue}\bf [$\rightarrow$ #1]}}

\begin{document}
%======================================================
 \title{Polaritonic states in a dielectric nanoguide: \\ localization and strong coupling}
 \author{Harald R. Haakh }%
 \email{harald.haakh@mpl.mpg.de }
\affiliation{Max Planck Institute for the Science of Light, G\"unther-Scharowsky-Stra{\ss}e 1, D-91058 Erlangen, Germany.}
 \author{Sanli Faez}%
\affiliation{Max Planck Institute for the Science of Light, G\"unther-Scharowsky-Stra{\ss}e 1, D-91058 Erlangen, Germany.}
\affiliation{Present Address: Debye Institute for Nanomaterials Science and Center for Extreme Matter and Emergent Phenomena, Utrecht University, Princetonplein 5, 3584CC, Utrecht, The Netherlands}
\author{Vahid Sandoghdar}
\affiliation{Max Planck Institute for the Science of Light, G\"unther-Scharowsky-Stra{\ss}e 1, D-91058 Erlangen, Germany.}
\affiliation{Friedrich Alexander University Erlangen-Nuremberg, D-91058 Erlangen, Germany}

\date{28 October 2015}
%\maketitle%revtex 3

%========================================
\begin{abstract}

Propagation of light through dielectrics lies at the heart of optics. However, this ubiquitous process is commonly described using phenomenological dielectric function $\varepsilon$ and magnetic permeability $\mu$, i.e. without addressing the quantum graininess of the dielectric matter. Here, we present a theoretical study where we consider a one-dimensional ensemble of atoms in a subwavelength waveguide (nanoguide) as fundamental building blocks of a model dielectric. By exploring the roles of atom-waveguide coupling efficiency, density, disorder and dephasing, we establish connections among various features of polaritonic light-matter states such as localization, super and subradiance, and strong coupling. In particular, we show that coherent multiple scattering of light among atoms that are coupled via a single propagating mode can gives rise to Rabi splitting. These results provide important insight into the underlying physics of strong coupling reported by recent room-temperature experiments with microcavities and surface plasmons. 
\end{abstract}

%========================================

\maketitle
%========================================
\section{Introduction}
%========================================

%
Although nineteenth century physics had difficulties with the concept of wave propagation without a supporting medium, light propagation in vacuum is well accepted in modern physics. On the contrary, a quantum mechanical first-principle description of light traveling through a material medium turns out to be subtle. Indeed, the derivations of dielectric function and magnetic permeability found in textbooks rely on effective fields and do not consider the quantum optical role of each constituent atom of the medium~\cite{Feynman1969, Jackson1999}. 

Experimental advances of the past two decades make it possible to explore dielectric media that are highly doped with molecules, quantum dots, color centers, or ions~\cite{Moerner:99}, while registering the position and transition frequencies of the individual emitters with nanometer spatial and MHz spectral resolution. Also, cold atoms can be prepared at relatively high densities. It is, thus, timely to revisit the problem of light propagation in dense media. In this Letter, we examine theoretically the interaction of a propagating light field and a collection of atoms. In particular, we consider a subwavelength waveguide (nanoguide), where each atom can reach a substantial coupling to the incoming field. 

%##################
\begin{figure}[t]
\centering 
 \includegraphics[width=\columnwidth]{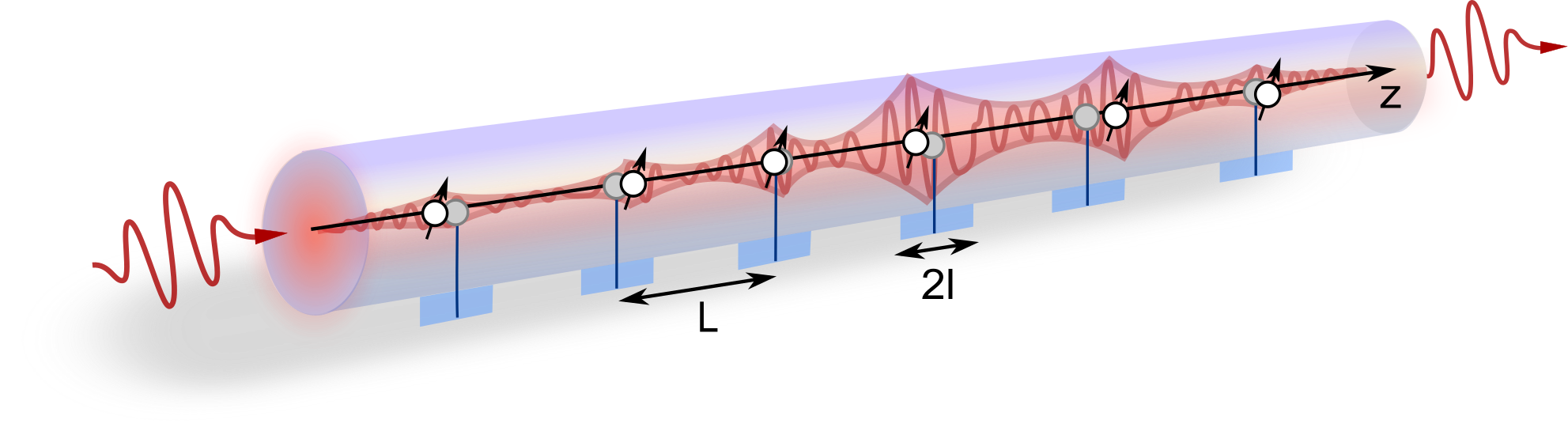}
 \caption{Scheme of a disordered many-body system inside a nanoguide. The emitters can be displaced within a region $\pm l$ (blue shading) around a regular lattice of spacing $L$ (gray dots). The red envelope indicates an example of a state extended over many emitters.
 }
\label{fig:system}
\end{figure}
%##################
Recently, we pointed out that one could achieve an efficient coupling of a single emitter to a propagating photon via tight focusing  and good mode matching~\cite{Wrigge2008, Zumofen2008, Tey2008}. However, coupling a light beam with a three-dimensional ensemble of emitters extended over regions beyond a wavelength remains a challenge. A tightly focused beam couples well only to a small number of atoms in the focus region, whereas an extended beam such as a plane wave has the right mode for the ensemble but couples weakly to the individual atoms. A nanoguide depicted in Fig.\,\ref{fig:system} presents a clever and efficient route for coupling a large number of material particles to a single mode of an optical field ~\cite{Balykin2004, Shen2005}. The nanoguide acts as an antenna that converts the dipolar radiation pattern of an atom and matches it to the fundamental spatial mode of the dielectric waveguide. Our findings show that this simple configuration is very rich in physics, bridging phenomena such as multiple scattering and disorder~\cite{Foldy1945, Lax1951,Lagendijk1996, Kramer1993, John1991, Lagendijk2009} to strong coupling and Rabi splittings in ensembles \citep{Keeling2007, Torma2015}.

Nanoguides have been used to couple light to single quantum dots~\cite{Akimov2007, Lund-Hansen2008, Yalla2012}, atoms~\cite{Vetsch2010, Goban2014}, color centers~\cite{Kolesov:09}, molecules~\cite{Stiebeiner2009, Faez2014a} and superconducting qubits~\cite{Astafiev2010, Loo2013}. On the theoretical side, the interaction of light and quantum emitters in a nanoguide has been studied for obtaining Bragg cavities~\cite{Chang2012}, coupling two atoms via a photonic channel~\cite{Gonzalez-Tudela2011} or realizing photon-photon interaction via an atom~\cite{Shen2007, Chang2012}. In what follows we present our theoretical formalism and show how the waveguide-mediated interactions affect the formation of collective and polaritonic modes depending on density, disorder, nanoguide coupling efficiency and homogenous broadening. 

%========================================
\section{The formalism}
%========================================
We investigate the propagation of light through a large number ($N$) of point-like two-level emitters (``atoms'') placed in a dielectric nanoguide (Fig.\,\ref{fig:system}). To facilitate the discussion, we assign each atom to site $n$ of a regular lattice with spacing $L$ along the waveguide axis and allow for longitudinal disorder by introducting a displacement $|l_n|$ for each site, which is smaller than or equal to a maximal displacement $l$.
All atoms have the same resonance frequency and radial dipole moment orientation and are characterized by the semiclassical polarizability $\alpha(\Delta) \approx \frac{1
%|\mu^2| / \hbar
}{ \Delta - \imath \Gamma/2}$, where $\Delta=\omega_{\rm A}-\omega_{\rm L}$ represents the  detuning between the frequencies of the atomic resonance ($\omega_{\rm A}$) and the driving light field ($\omega_{\rm L}$).
The natural linewidth $\Gamma = 2 \Im[G(0)]$ includes the radiation reaction \cite{Vries1998}  in the nanoguide  through the dyadic Green function  $G(z)$ \cite{Li2000}.
The coupling between an atom and the optical mode of the nanoguide is characterized by the parameter $\beta=\Gamma_{\rm ng}/(\Gamma_{\rm ng}+\Gamma_{\rm out})$, where $\Gamma_{\rm ng}$ and $\Gamma_{\rm out}$ denote the single-emitter emission rates into the nanoguide mode and the outside world, respectively. 

In the waveguide, a single dipole $d_n$  generates a field \mbox{$E_n^{\rm dip}(z)=G(z-z_n)d_n$} \cite{Li2000} so that the dipole moment at each site induced by an external field $E^{\rm ext} (z)$ and by other sites is given by \cite{Foldy1945, Lax1951}  (cf. Supplemental Material)
	\begin{equation}
	\label{eq:field}
	d_n \!=\! \alpha_n \biggl[E^{\rm ext} (z_n) + \sum_{m\ne n}E^{\rm dip}_m(z_n)
	% G(|z_m - z_n|) 
	%d_m
	\biggr]
	 =\sum_{m}\mathcal{A}_{mn} E^{\rm ext}_m~.
	\end{equation}
Equation\,\eqref{eq:field} gives the full field distribution in the multiply scattering system. $\mathcal{A}$ acts as a collective polarizability and is obtained by the explicit inversion of the nonhermitian matrix \mbox{$\mathcal{A}^{-1}_{mn} = \delta_{mn}/\alpha_n - G(z_n - z_m) (1-\delta_{mn})$}. The real and imaginary parts of the complex eigenvalues \mbox{$(\delta - \imath \gamma/2)$} give the shift of the collective resonance from the single emitter one and its linewidth, respectively. The $N$-component mode functions $\Psi = (d_1, d_2, \dots, d_N)/\sqrt{\sum_n |d_n|^2}$ describe normalized self-sustained dipole-moment distributions along the chain.
For each collective (polariton) mode function, the participation number \mbox{$p = \sum_{n=1}^N|\Psi_n|^4$} counts the optically active emitters \mbox{$1\le p \le N$} and quantifies the extension of the state \cite{Kramer1993}. 

%========================================
\section{Collective spectra in far-field coupled chains}
\subsection{The ordered case} 
%========================================
We first consider ordered chains ($l=0$) with  large spacings  $L>\lambda$ so that near-field effects are  negligible. Here, interactions are mediated by the guided mode with a wavenumber $k=2\pi/\lambda$ and \mbox{$G(z) \approx \imath \beta \Im[G(0)] e^{\imath k |z|}$} \cite{Gonzalez-Tudela2011,  Haakh2015}.
For even $N$ and a chain under Bragg or anti-Bragg conditions (odd or even $2L/\lambda\in\mathbbm{N}$, respectively), $\mathcal{A}^{-1}$ can be exactly diagonalized by a Fourier ansatz (cf. Supplemental Material). We find a single superradiant eigenstate with the linewidth 
\begin{equation}
\gamma_+ = [1+\beta(N-1)]\Gamma
\end{equation}
and a mode function \mbox{$\Psi = (1,1,\dots)/\sqrt{N}$} (even $2L/\lambda$)  or  \mbox{$\Psi = (1,-1,\dots)/\sqrt{N}$} (odd $2L/\lambda$) that is perfectly matched to the waveguide mode.
Scattering from this state alone yields efficient (Bragg) reflection and ohmic transmission $T\propto N^{-2}$ as $N\to \infty$ \cite{Chang2012}, which can be understood as coherent extinction by a collective superdipole.
The remaining $N-1$ eigenstates are degenerate and subradiant with
\begin{equation}
\label{eq:gammaminus}
\gamma_- =(1-\beta)\Gamma~. 
\end{equation}
The corresponding mode functions are periodic but do not match the guided mode and cannot be excited through the nanoguide. Since the states provide a complete basis with either minimal or maximal mode matching to the waveguide field, $\gamma_-$ and $\gamma_+$ describe general bounds for the decay rates in far-field coupled chains.

\begin{figure}[t!]
\includegraphics[width=.9\columnwidth]{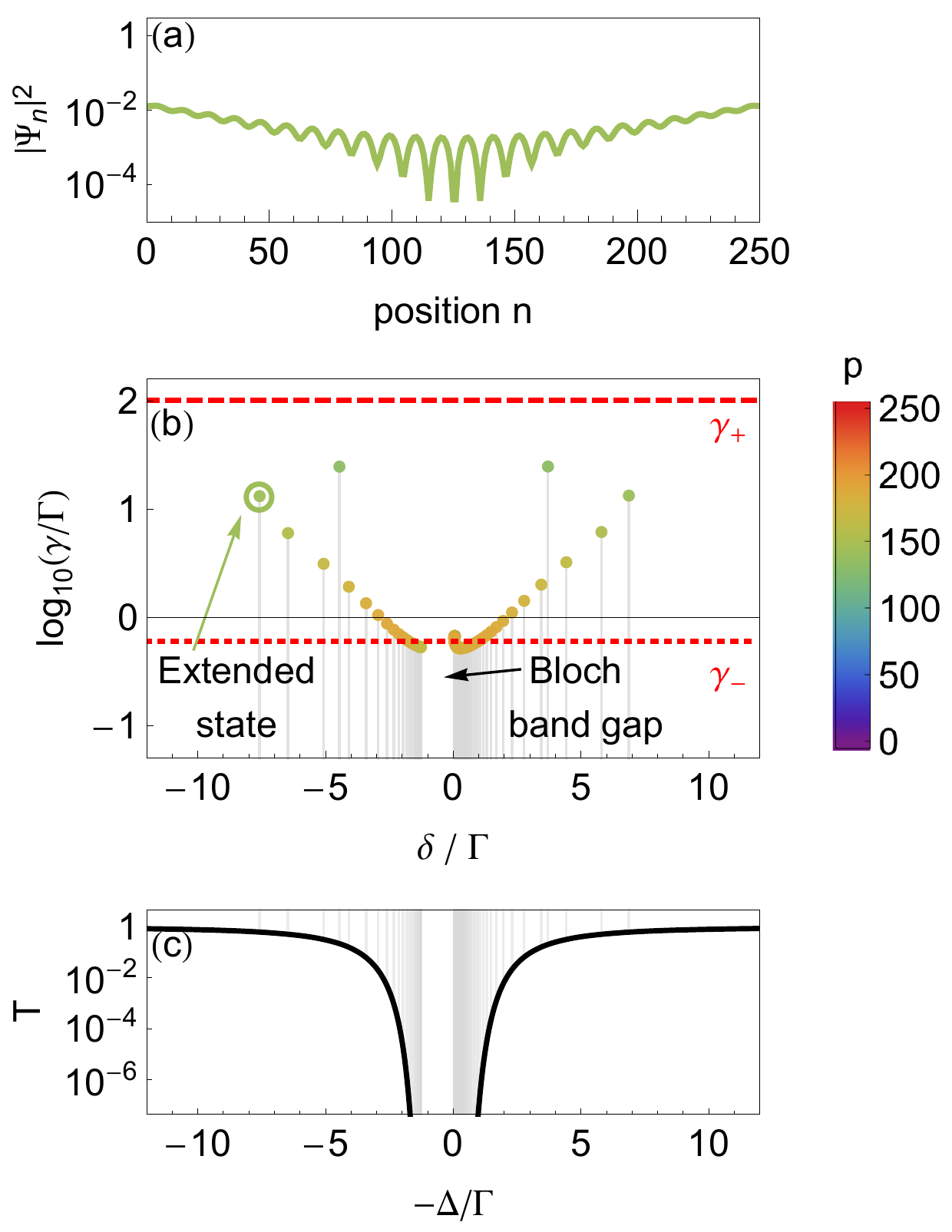}
 \caption{In this figure we consider an ordered far-field coupled chain of $N=250$ emitters for $L=2.75\lambda, l=0, \beta=0.4$. 
 a) Exemplary mode function of an extended state.
  b) The 250 eigenvalues forming the polariton spectrum plotted as a function of their real ($\delta$; horizontal) and imaginary ($\gamma$; vertical) parts. The color coding of each resonance gives the participation number of the respective mode function. The state marked by the green circle corresponds to the extended  mode function displayed in (a). The decay rates $\gamma_\pm$ provide upper and lower limits to the polaritonic linewidths.
 c) Transmitted power through the guided mode as a function of the frequency detuning of the driving field from atomic resonance.
 }
 \label{fig:specExt}
\end{figure}

At different periodicities, extended Bloch-like states arise. Figure\,\ref{fig:specExt}(a) shows the squared amplitude of exemplary extended mode function ($p = 142$) arising in a lattice of $N=250$ emitters with $L=2.75 \lambda, l=0$, and $\beta=0.4$.
In Fig.\,\ref{fig:specExt}(b), we identify the complex eigenvalues of the emitters on the real (frequency shift) and imaginary (linewidth) axes, revealing a band gap around the single-emitter resonance ($\delta=0$). 
We see that subradiant extended modes with decay rates $\gamma_- \le \gamma < \Gamma$ gather at the band edges, and all collective modes involve $p \approx 100\dots 200$ emitters as indicated by the color-coded participation number assigned to each resonance.

To examine the collective response of the ensemble on an incoming photon, in Fig.\,\ref{fig:specExt}(c) we display the transmission through the nanoguide as a function of frequency. As intuitively expected, the transmission drops near the band gap, in fact it drops faster than exponentially in $\Delta$. The dependence of the transmission remains, however, exponential in terms of the number of emitters (not shown).
%###########################
%###########################

%###########################
\begin{figure}[t!]
\vspace{1ex}
\includegraphics[width=.9 \columnwidth]{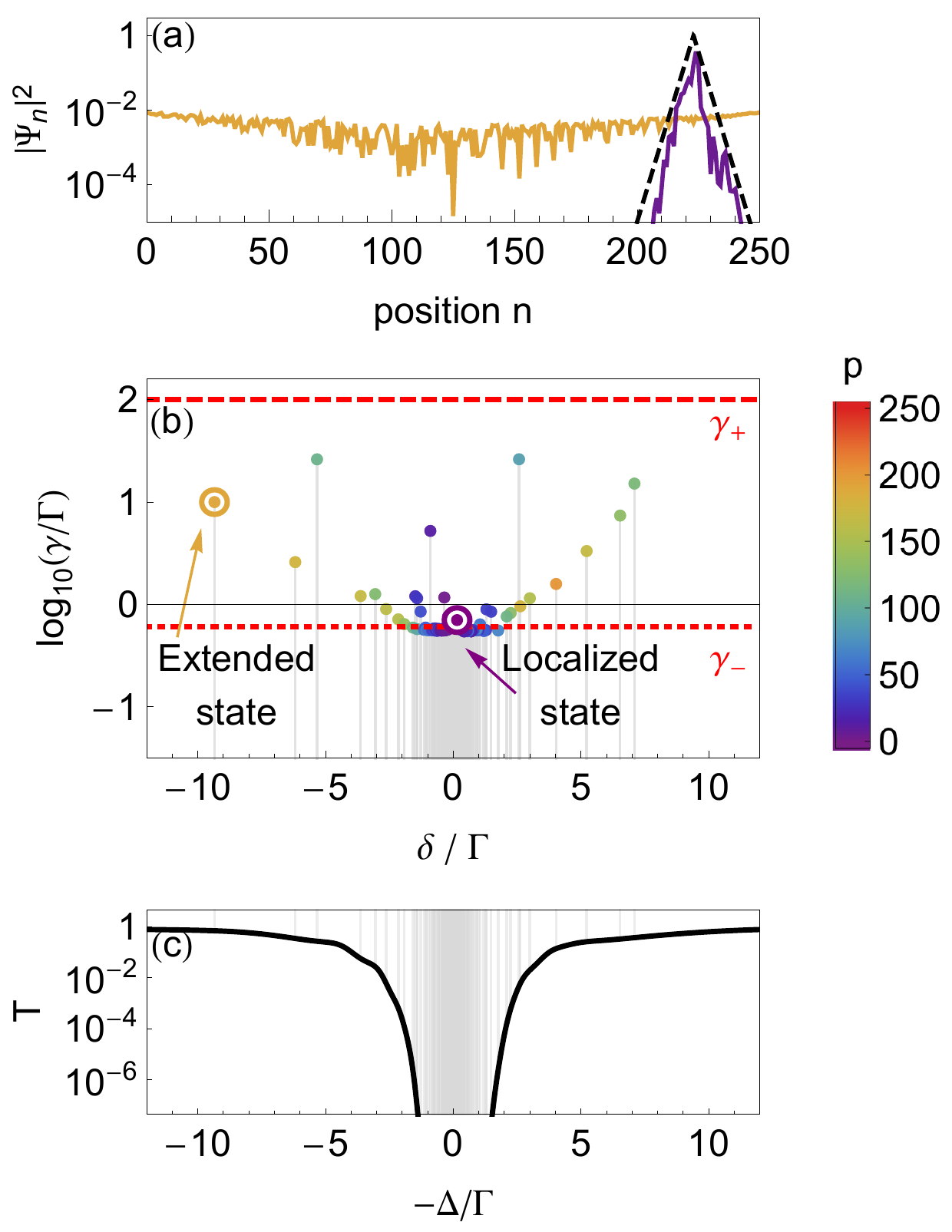}
 \caption{In this figure we consider an disordered far-field coupled chain of $N=250$ emitters for $L=2.75\lambda, l=\lambda/2, \beta=0.4$. 
 a) Exemplary mode functions of an extended (orange) and  localized state (purple). The black dashed curve represents an exponential envelope.
 b) The 250 eigenvalues forming the polariton spectrum plotted as a function of their real ($\delta$; horizontal) and imaginary ($\gamma$; vertical) parts. The color coding of each resonance gives the participation number of the respective mode function. The states marked by orange and purple circles corresponds to the extended and localized mode functions displayed in (a), respectively. Again, the decay rates $\gamma_\pm$ provide upper and lower limits to the polaritonic linewidths. Here, we see that subradiant localized states ($\gamma<\Gamma, p\ll N$) fill the band gap.
 c) Transmitted power through the guided mode as a function of the frequency detuning of the driving field from atomic resonance.
 }
 \label{fig:specLoc}
\end{figure}
%###########################

%======================
\subsection{The disordered case} 
%======================
The more general case of a dielectric medium concerns a glassy environment with an intrinsic degree of disorder. In conventional dielectrics, the scattering cross sections of the individual microscopic Huygens entities are very small so that multiple scattering does not play an important role. In our case, each atom in the nanoguide has a substantial influence on an incoming photon so that one might expect light localization effects \cite{John1991, Lagendijk2009}. In fact, it is known that even a small degree of disorder gives rise to localization in truely one-dimensional systems~\cite{Kramer1993, Bliokh2006}. By considering finite values of $\beta$, we now investigate the effect of scattering loss on localization.

Let us consider a chain of lattice spacing $L=2.75 \lambda$ and $\beta =0.4$ as before but set $l = \lambda/2$ to provide full phase randomization and disorder within the frame of guided far-field coupling. In Fig.\,\ref{fig:specLoc}(a) we present two exemplary resulting mode functions (orange and purple). The orange curve displays a super-radiant mode that preserves the extended character of its ordered counterpart discussed in Fig.\,\ref{fig:specExt}(a), while the purple curve shows a characteristic exponential envelope that corresponds to localized states. These states are marked in Fig.\,\ref{fig:specLoc}(b), which plots the complex eigenvalues of the various modes.

The resemblance of the eigenvalue distribution in the superradiant sector of Fig.\,\ref{fig:specLoc}(b) reveals a clear similarity between the extended wavefunctions of the ordered and disordered chains. In contrast, localized states gather in the subradiant sector where participation numbers $p\ll N$ indicate the confinement of the excitation to a limited region of the chain. We point out that, nevertheless, the decay rates remain confined between the two dashed red curves $\gamma_\pm$ that limit far-field coupled chains. Subradiant states move into and eventually close the band gap as $l$ increases, but they do not significantly increase the transmission in the band gap region [see Fig.\,\ref{fig:specLoc}(c)] because their exponential mode shape provides poor mode-matching to the waveguide mode and favors scattering out of the mode.

Subradiance, reduced participation number, and exponential mode function envelopes indicate the persistence of longitudinally localized polaritons despite limited $\beta$. However, the dependence of $\gamma_-$ on the coupling efficiency in Eq.\,\eqref{eq:gammaminus} shows that transverse loss to the 3D continuum limits the polariton lifetime independently of the number of scatterers. As a direct consequence, the Thouless criterion of Anderson localization \cite{Abrahams1979}, which requires ever stronger line-narrowing with increasing particle number remains indecisive unless $\beta \to 1$. This differs fundamentally from 3D arrangements of emitters \citep{Jonckheere2000, Skipetrov2014}.

%========================================
\section{Near-field coupling and spectral selectivity }
%========================================
%
%###########################
\begin{figure}[t!]
\includegraphics[width=.9\columnwidth]{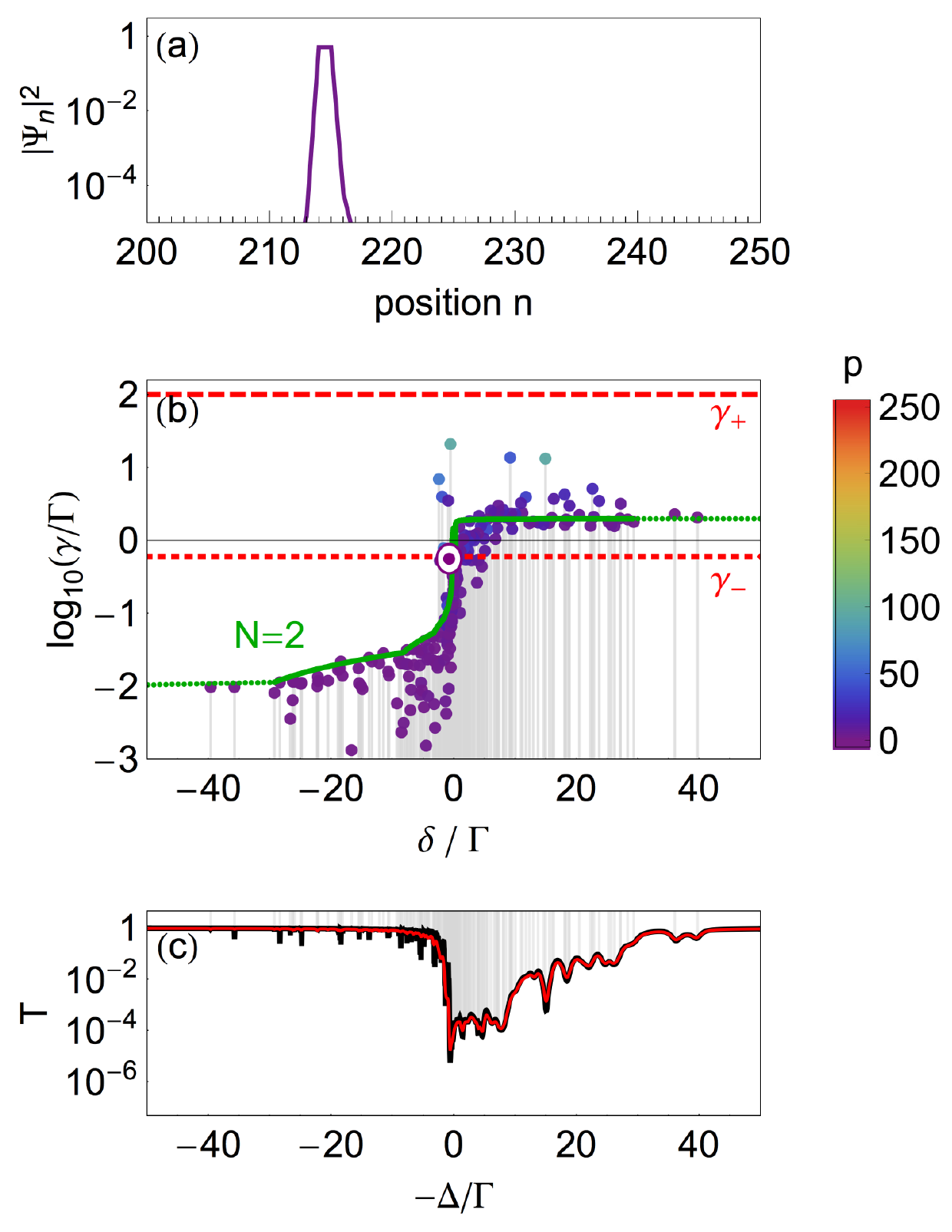}
 \caption{Here, we consider a disordered near-field coupled chain of atoms with $L=\lambda/2 \pi, l=0.49L, N=250, \beta=0.4$
 a) Exemplary mode function of a two-body resonance. b) The 250 eigenvalues forming the polariton spectrum plotted as a function of their real ($\delta$; horizontal) and imaginary ($\gamma$; vertical) parts. The color coding of each resonance gives the participation number of the respective mode function. The state marked by the purple circle corresponds to the state shown in (a). The green curve displays the expected results for the eigenvalues of a pair states ($p = 2$) on a minimal chain of $N=2$ emitters.
 c) Strongly asymmetric transmission spectrum for the near-field coupled disordered chain (black curve). The red curve includes additional nonradiative losses  $\Gamma_{\rm hom} = 0.25 \Gamma$ which smoothens the narrow lines.%
 }
 \label{fig:specTBS}
\end{figure}
%###########################
%

At high densities  ($L\ll\lambda$) and/or strong disorder \mbox{($L-l\ll\lambda$)} neighboring atoms are coupled by a near-field potential \mbox{$G(z)\propto-1/z^3$} \citep{Rusek2000, Skipetrov2014,Caze2010}. Here, the linear arrangement of emitters in the nanoguide results in next-neighbor interactions that dominate over the coupling to any other emitter, leading to the fractionalization of the nanoguide to chainlets of neighboring pairs even at moderate disorder.
Figure\,\ref{fig:specTBS}(a) displays the pair character of the eigenstates in an exemplary mode function for a chain of $N=250$, $L=\lambda/2\pi$, and $l=0.49 L$.
In this regime, the previously obtained general limit \mbox{$\gamma_-\le\gamma\le\gamma_+$} no longer applies, and the polariton spectrum shows a significant number of strongly subradiant states with decay rates several orders of magnitude below $\gamma_-$ [see Fig.\,\ref{fig:specTBS}(b)]. These sharp resonances generate distinct features in the transmission spectrum shown in Fig.\,\ref{fig:specTBS}(c). 

The subradiant two-body states do not overlap spectrally, and their odd wavefunctions [$\Psi_- \approx (0,\dots, 0, 1,-1, 0, \dots,0)/\sqrt2$] are hardly driven by the waveguide mode, which varies slowly across the pair. As a result, we obtain a high transmission at $-\Delta<0$. In contrast, the superradiant resonances spectrally overlap and provide strong scattering and losses from the waveguide at $-\Delta>0$ (cf. Supplemental Material for details). This induces the steep change of more than 4 orders of magnitude in the transmittivity over few atomic linewidths. The coincidence between the transmission dips and polariton resonances at $-\Delta = \delta$ is underlined by the vertical projections in Fig.\,\ref{fig:specTBS}(b, c). 

The dominant purple hues in Fig.\,\ref{fig:specTBS}(b) show that almost all polariton modes, including the most subradiant ones near $\delta \approx 0$, are strongly confined to $p<20$ emitters. In fact, it turns out that the participation number distribution is strongly peaked at $p=2$. This comes along with a strongly asymmetric shape of the spectrum in the complex plane, where a significant fraction of the states gather near the green curve , which is obtained for a chain of $N=2$ emitters found at different spatial separations, leading to one super- and one subradiant state \citep{Rusek2000, Skipetrov2014}.

While in the guided far field, $G(z)$ could acquire any complex phase, providing both super- and subradiant states at positive or negative values of $\delta$, the phase is fixed in near-field coupling. As a result, $\gamma$  and $\delta$ become correlated and the polaritonic states gather in red-detuned subradiant and blue-detuned superradiant sectors in the complex plane. To illustrate this situation for solid-state systems, in Fig.\,\ref{fig:specTBS}(c) we consider moderate dephasing by replacing $\Gamma\to \Gamma_0+ \Gamma_{\rm deph}$ in the denominator of the polarizability, where now $\Gamma_0$ is the natural linewidth. $\Gamma_{\rm deph}$ provides a lower limit to the collective linewidths and, as displayed by the red curve in Fig.\,\ref{fig:specTBS}(c), the sharp features of the system are smoothed out.  

%========================================
\section{Resonance-free polaritonic strong coupling}
%========================================
Above we have studied the basic influence of order, disorder and near-field coupling between the atomic consituents of a dielectric nanoguide on its optical response. We now show that for large enough densities, the collective coherent interaction of the atoms with the single guided mode of the waveguide leads the hybrizidation between the atomic ensemble and the nanoguide mode, an effect that is usually known in coupled oscillators.

Rabi splitting in the solid state was realized soon after the first gas-phase studies of strong coupling between an atom and a high-$Q$ cavity~\cite{Haroche-book:06}. To achieve this, scientists investigated the reflectivity and transmission of light from quantum wells embedded in microcavities at cryogenic temperatures~\cite{Weisbuch:92}. This pioneering work has triggered similar activities on ensembles of organic molecules coupled to low-$Q$ microcavities at room temperature~\cite{Lidzey:98, Kena-Cohen:08, Schwartz2012}, where a large number of emitters show collective strong coupling despite substantial homogeneous and inhomogeneous spectral broadenings. Interestingly, similar mode splitting has also been reported for an ensemble of molecules coupled to surface plasmons~\cite{Hakala2009}. The physical picture describing these experiments has been the coupling of two pendula with well-defined resonances, whereby the collection of emitters is modelled with an effective polarizability~\cite{Torma2015}.
We now show that although the nanoguide considered in this work only provides a propagating mode without any resonance, it can nevertheless hybridize with atoms at sufficiently high densities. In particular, our treatment shows how the collective effect of emitters emerges from multiple scattering and interference of the light field confined within the nanoguide mode. 

%
%########################### 
\begin{figure}[t!]
\includegraphics[width=.99\columnwidth]{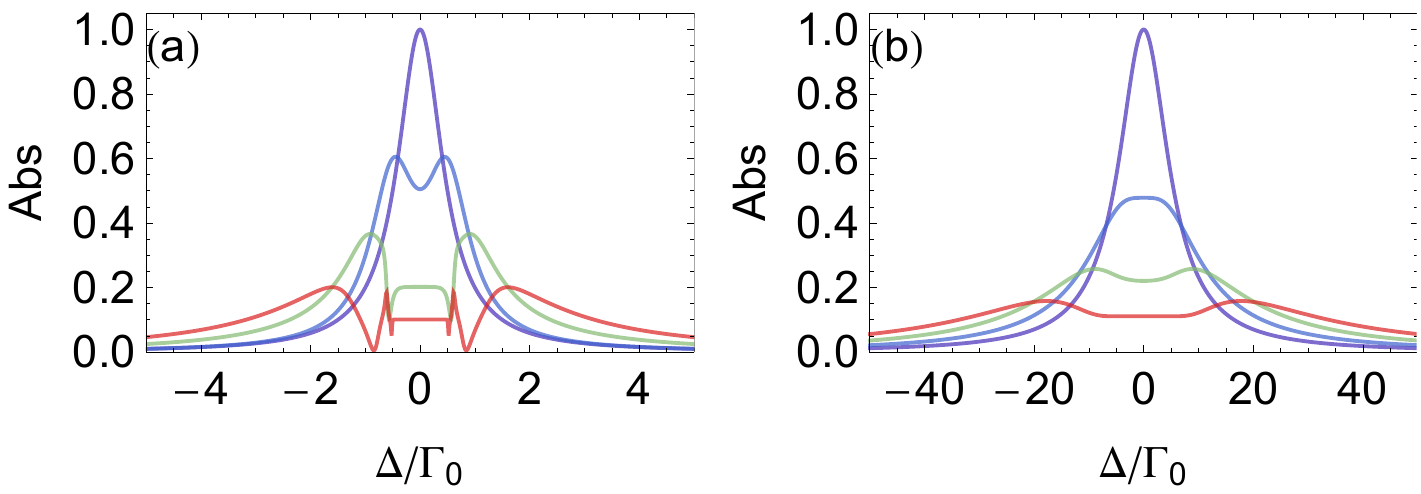}
\caption{Collective strong coupling for distant emitters. a) Absorbtion by an ordered chain of $L = 2.75\lambda$ in a near-ideal waveguide $\beta = 0.99$ for $N=1,2,5,10$ (purple to red) emitters. The signal is normalized to the absorption by $N$ uncoupled resonant emitters. The red curve shows narrow peaks near $\Delta \approx 0$ for subradiant modes.
b) Same as in (a) but for chains of $N=10,20,50,100$ (purple to red) emitters in the case of a homogeneous broadening with $\Gamma_{\rm deph} = 10 \Gamma_0$.}
\label{fig:splitting}
\end{figure}
%###########################

To gain intuition into the underlying physics of polaritonic Rabi splitting, we first consider the simplest case of two individual atoms coupled via a waveguide~\cite{Rephaeli2011, Martin-Cano2010}. Since a single atom with $\beta=1$ can act as a perfect mirror for a photon entering the nanoguide~\cite{Zumofen2008,Shen2009}, two atoms placed at a large distance can form an optical resonator. Unlike in a usual Fabry-P\'erot cavity, the ``mirrors'' are influenced by each other's presence. The delocalized collective states in this case are described by mode functions $\Psi_{1,2} = (1, \pm 1)/\sqrt{2}$ and complex resonances \mbox{$\delta_{1,2}- \imath \gamma_{1,2}/2 = 1/\alpha \pm G(z_1 - z_2)$}. Thus, the splitting $|\delta_1 - \delta_2|\le \beta \Gamma$ is maximized at $L = (2 m+1) \lambda/4, m \in\mathbbm{N}$ for the case of far-field coupling.
 
Figure\,\ref{fig:splitting}(a) illustrates the frequency-dependent absorption $\propto\sum_{m}\Im[d_m E^{\rm ext*}(z_m)]$ (normalized to the case of uncoupled emitters on resonance) in chains of few atoms spaced at $L = 2.75\lambda$ in a waveguide of $\beta = 0.99$. The splitting for two emitters is clearly visible. Increasing the number of emitters leads to multiple scattering and formation of many collective modes as discussed in the previous sections.  It is also evident that the splitting between the absorption maxima grows.
To investigate the effect of coherence and large homogeneous broadenings encountered in room-temperature solid-state systems, in Fig.\,\ref{fig:splitting}(b) we allow for a ten-fold homogeneous broadening of $\Gamma_{\rm deph}=10\Gamma_0$. We find that spectral splitting still occurs, but it requires a higher number of emitters to compensate for a reduced coupling efficiency, which in turn causes less efficient hybridization among emitters. Furthermore, we note that the faster spectral variations are washed out.

%########################### 
\begin{figure}[t!]
\includegraphics[width=.99\columnwidth]{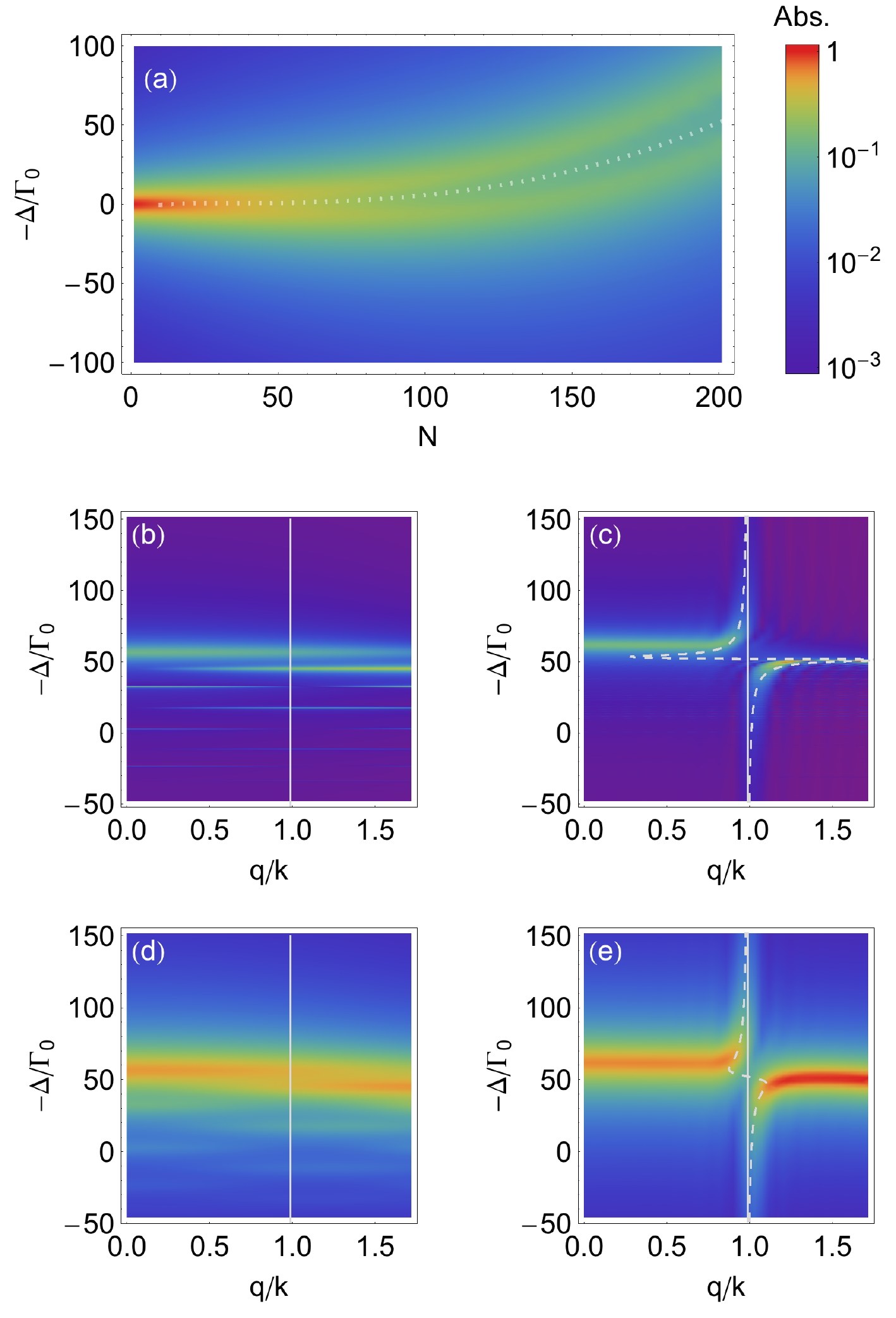}
\caption{Collective strong coupling in dense chains. a) Absorbed power as a function of the frequency detuning of the driving field as a function of an increasing number of emitters. The waveguide is taken to have $l=0$ and $\beta=0.56$ filled to a constant length of $N L = 7.7 \lambda$. The atoms are assumed to be homogeneously broadened with $\Gamma_{\rm hom}=10\Gamma_0$. The white dotted curve plots the expected near-field shift. \break
  b - e) Dispersion relations obtained by evaluating the absorbed power of life-time limited atoms as a function of the wavenumber $q$ of the driving field $E^{\rm ext}(z) = E_0 e^{i q z}$ and its frequency detuning. b) We consider a dense ordered chain of $L=0.04\lambda, \Gamma_{\rm deph} = \Gamma_0$ for $N=10$, where no emitter-waveguide hybridization is visible. c) $N=200$ emitters, where hybridized excitations approach the dispersion relation in the background waveguide.  d, e) Same as (b, c) but for $\Gamma_{\rm deph} = 10\Gamma_0$. The dashed white curves give the analytic result of a mean-field approach in each case.}
 \label{fig:hybridization}  
  \end{figure}
%########################### 

Next, we increase the density of emitters in a waveguide of fixed length $NL=7.7 \lambda$ and finite $\beta =0.56$, hosting emitters with additional broadening  $\Gamma_{\rm deph}=10\Gamma_0$. Figure\,\ref{fig:hybridization}(a) displays the absorption of light propagating through the nanoguide medium as a function of $N$ and frequency detuning $\Delta$ from a single-emitter resonance. We see that one can indeed arrive at a large frequency splitting for higher densities (or equivalently $N$). The general shift that accompanies the splitting for larger $N$ is caused by near-field interactions among neighboring atoms. This effect can be approximated by $-\Re[G(L)]$ plotted by the white dotted curve. The important result here is the observation of a spectral splitting in a dense ensemble of atoms simply because their emission is coupled efficiently enough to a nanoguide to cause hybridization between its mode and their collective emitter resonances.
To this end, one can also identify the main role of a microcavity \citep{Weisbuch:92, Keeling2007} or surface plasmon \citep{Gonzalez2013, Delga2014, Torma2015} in offering an efficient single mode for efficient multiple scattering of many emitters. In this picture, the spectral or angular resonances of such structures are not of fundamental importance.  

A decisive indicator of collective hybridization with the guided mode is a repulsive splitting in the dispersion relation of the system at hand~\cite{Torma2015}.
In Fig.\,\ref{fig:hybridization}(b,c) we plot the absorption of a nanoguide containing an atomic chain as a function of the frequency detuning $\Delta$ of the driving field and its wavenumber (periodicity) $q$.
Driving of the nanoguide with different wavevectors can be experimentally achieved in a Kretschmann configuration. 
Fig.\,\ref{fig:hybridization}(b) shows that multiple peaks at a given value of $q$ are still visible in a sample as small as $N=10$ while there is no clear avoided crossing. In longer chains, on the other hand, the subpeaks are suppressed and the absorption maximum splits into two branches that approach the dispersion curve $k(\Delta)$ of the empty waveguide [see white solid curve in Fig.\,\ref{fig:hybridization}(c) for $N=200$]. Dispersion of the latter is negligible since $\omega_A/\Gamma_0 \approx 10^6$. Figs.\,\ref{fig:hybridization}(d,e) show that all these features persist when a broadening $\Gamma_{\rm deph}=10\Gamma_0$ is assumed.

Finally, we examine the observed normal mode splitting in the context of an effective medium response~\cite{Torma2015}. When $L\ll\lambda\ll NL$, e.g. for the cases of Fig.\,\ref{fig:hybridization}(c,e), we can apply a mean-field approximation \cite{Parola2014} and restrict the impact of near-field coupling to the above global shift, to obtain an analytical expression of the wave number inside the filled region (see Supplemental material for details). To lowest order in the polarizability, the system response at a wave number $q=\tilde n \omega/c$ fulfills a generalized Clausius-Mossotti relation
\begin{align}
\frac{\tilde n-1}{\tilde n+2}\approx \frac{\tilde \alpha}{3 \varepsilon_0}\times\frac{1}{L  A_{\rm eff} n_{\rm eff} }~.
\end{align}
We note that here $A_{\rm eff}$ and $n_{\rm eff}$ denote the effective mode cross-section and refractive index of the empty waveguide mode, and the polarizability $\tilde\alpha$ is given in SI-units. The white dashed curves in Fig.\,\ref{fig:hybridization}(b,c), which plots $\Re[q(\Delta)]$, reveal a very good agreement between this analytical phenomenological result and the numerical calculations when sufficiently large numbers of emitters are considered.

%========================================
\section{Discussion}
%========================================

In this work, we have developed a semiclassical theoretical platform for examining the interaction of light with a collection of atoms in a one-dimensional subwavelength waveguide, where the atomic emission is efficiently coupled to the waveguide mode. Our approach is based on the explicit solution of the multiple-scattering problem in terms of collective response functions, using the exact dipole emission pattern in a cylindrical nanoguide and a semiclassical polarizability. This formalism is in excellent agreement with a quantum-optical master equation for the case of weak driving fields and can be readily extended to multilevel atoms~\cite{Witthaut2010} by using a modified polarizability. 

We based our treatment on a one-dimensional lattice of atoms that could be randomly displaced by $l$, up to one lattice constant $L$. However, the calculations can also be extended to $l > L/2$ to allow for other disorder models \cite{Kramer1993} and clusters beyond pairs to consider excluded volumes and possible self-organization \cite{Chang2013}. In addition to spatial disorder, the model could also be extended to include inhomogeneous spectral distributions as commonly encountered in solid-state systems. 

Although an ideal scenario of a 1D nanoguide strives for an emitter-waveguide coupling efficiency $\beta \rightarrow 1$, real experimental systems will have serious difficulties to reach this limit. We have, thus, chosen to work with moderate values of $\beta$ and have shown that while low values of $\beta$ degrade the hybridization between the emitters, significant collective coupling between the emitters and the waveguide mode remains feasible and can be compensated by adding more emitters. In this sense, $\beta$  plays the same role as in the performance of microcavities.

An important message of our work is to emphasize the need for considering the near-field dipole-dipole coupling that results from subwavelength inter-atomic spacings. This situation arises both for very dense regular chains and for large degrees of disorder as known from glassy dielectrics. In either case, a strong near-field coupling generates spectrally selective narrow features in the optical response.

Our nanoscopic model of multiple scattering describes mutual hybridization while including the effect of decoherence and linewidth broadening. We have seen that each realization of the ensemble may contain many spectral features that arise from near-field interactions or from disorder and state localization. By averaging over a large number of such individual configurations, one can arrive at an effective optical response for the corresponding one-dimensional dielectric medium. In fact, we demonstrated that a mean-field limit asymptotically agrees with our explicit solution of the scattering problem. This provides a link between the present work and effective medium descriptions of light propagation through a resonant medium. 

%========================================
\section{Conclusion}
%========================================

Experimental activities on one-dimensional waveguide systems have undergone a rapid growth in the past few years in a range of materials covering gas-phase atoms~\cite{Vetsch2010, Goban2014}, semiconductor quantum dots~\cite{Akimov2007, Lund-Hansen2008, Yalla2012}, molecules~\cite{Faez2014a} and even superconducting qubits~\cite{Astafiev2010, Loo2013}. Our theoretical results show that these systems can be exploited to study a wealth of interesting collective phenomena in a compact and scalable fashion. In particular, this platform provides a very promising avenue for exploring polaritonic physics \cite{Keeling2007} in a well-controlled environment. As a concrete example, we have demonstrated that it should be possible to achieve normal mode splitting in a nanoguide without involving structures that support resonances such as a cavity or a plasmonic structure.

%========================================
%========================================

\textbf{Acknowledgments}
We thank D. Mart\'in-Cano, I. Chremmos, and P. T\"urschmann for helpful comments. V.S. also acknowledges fruitful discussions with Jean-Jacques Greffet. This work was supported by an Alexander von Humboldt professorship, the Max Planck Society and the European Research Council (Advanced Grant SINGLEION).

%\small
\normalsize
	
%\bibliographystyle{aipnum4-1}
%\input{2015-10-21-bib.bbl}
%\bibliography{localization-extract}

%\bibliography{vahid,localization-letter}
%\bibliography{/Users/hhaakh/Documents/literatur/Bibliography/bibliography}

%

%\newpage
%~

%========================================
\appendix*
%========================================

\newpage

%========================================
\section*{Supplemental Material}
%========================================
\subsection{Waveguide Green function}
We consider a system of two-level emitters placed on the axis of a dielectric cylinder with radius $R=160 \dots 215 \unit{nm}$ and refractive index $n=1.5$ surrounded by vacuum. For simplicity, we only consider radial components of the polarizability.
The numerical simulations were performed for two-level emitters with transition at $760\unit{nm}$ and a natural linewidth of $\Gamma_0= 400 \unit{MHz}$, possibly modified by dephasing at a rate $\Gamma_{\rm deph}$. Imposing modified atomic units, the transition dipole matrix element is $|\mu| = \hbar = 1$, and the polarizability in the near-resonant approximation is $\alpha(\Delta) = \frac{1}{\Delta - \imath \Gamma/2}$, with $\Gamma = \Gamma_0+ \Gamma_{\rm deph}$.
The dyadic Green function in the cylinder is expressed in the scattering decomposition $G(z) = G^{\rm (free)}(z) + G^{\rm (sc)}(z)$. Closed forms of this quantity are available \cite{Li2000, Haakh2015} and were used for all calculations of complex eigenvalue spectra. In the single-mode regime, the  guided mode contribution with wavenumber $k= n_{\rm eff} \omega/c$ can be extracted from the resonances of the scattered term $G^{\rm(sc)}$ and yield the asymptotic far-field \cite{Gonzalez-Tudela2011, Haakh2015} 
\begin{align}
{G}(z) \approx \imath \beta \Im[{G}(0)] \exp(\imath k |z|).
\label{eq:GT1DFarfield}
\end{align}
The amplitude was expressed in terms of the local-field corrected single-emitter linewidth $\Gamma_0 = 2  \Im[G(0)]$ that includes the radiation reaction \cite{Vries1998}. 
For the above parameters, the coupling efficiency of a single emitter to the guided mode is $\beta = \lim_{z\to\infty}|{G}(z)|/ \Im [{G}(0)]\approx 0.4\dots0.56$. 
Different choices of the radius or material, e.g. IR-emitters inside a semiconductor cylinder (GaAs, $R=140$nm, transition at $1.2 \unit{\mu m}$) may reach $\beta > 0.9$ \cite{Claudon2010}. 
 Note that over the relevant narrow frequency band of a few hundred $\Gamma_0$, waveguide dispersion is neglible, as can be seen by the vertical white curve in Fig.\,\ref{fig:hybridization}(b,c).

\subsection{Bragg resonances}
We outline the calculation of the collective modes on a Bragg resonance $kL/ \pi \in \mathbbm{N}$ and assume even $N$. The latter restriction has, however, little impact at the large $N$ and finite $\beta<1$  considered here. In the far-field approximation, %
\begin{equation}
\mathcal{A}^{-1}_{mn} \approx \frac{\delta_{mn}}{\alpha_n} - \imath \beta  \frac{\Gamma}{2}\Im[{G}(0)]e^{\imath kL |m-n|} (1- \delta_{mn})
\label{eq:farfieldmatrix}
\end{equation}
 which is a circulant matrix and can be diagonalized by Fourier transformation. The eigenvalues are
\begin{align}
\delta_m - \imath \frac{\gamma_m}{2}
%&\approx \frac{1}{\alpha} - \imath \beta\frac{\Gamma}{2}  \sum_{j=1}^{N-1} e^{ \imath k L j} 
%e^{- 2 \imath \pi m j / N} \\
&\approx  \frac{1}{\alpha} - \imath \beta \frac{\Gamma}{2}  \frac{-1 + e^{- \imath  (2\pi m/N - k L) (N-1)}}{1- e^{  \imath  (\pi m / N - kL)}}~.
\label{eq:spectrumDicke_asymptote}
\end{align}
For even (odd) values of $kL/\pi$ a pole is encountered at $m=0$ ($m=N/2$). L'H\^{o}pital's rule then provides a single superradiant (bright) state, while the other  $N-1$ states are degenerate and subradiant (dark):  \myref{7/38}
\begin{align}
\delta_m - \imath \frac{\gamma_m}{2}&=  \frac{1}{\alpha} - \imath \beta \frac{\Gamma}{2} 
 \times \begin{cases}
 - 1 		&\text{dark}  \\
 N-1 	& \text{bright} 
 \end{cases}~.
\label{eq:spectrumDicke_asymptote2}
\end{align}
This agrees with a recent discussion that assumed $\beta=1$ \cite{Liao2015}.
The  mode functions are ($1\le j \le N$) 
$
\vec{\psi}_{m,j}
\approx e^{- 2 \pi \imath m (j-1)/ N}/{\sqrt{N}}.
$\label{eq:spectrumDicke_wf}
Only the single bright mode $\Psi_{+} = (1,\pm1,1,\pm 1,\dots)/\sqrt{N}$ matches the periodicity of the guided mode [$E^{\rm ext}(z_n) = E_0 e^{\imath k z_n}=E_0 \Psi_{{+},n}$], whereas the other modes have zero mode overlap.
To find the scattering matrix element, we neglect the dark modes and approximate 
\begin{align}
\mathcal{A}^{-1} \approx \Delta- \imath \frac{\Gamma}{2} [1 + \beta (N-1)] \Psi_{+}\otimes\Psi_{+}^T.\end{align}
In the collective basis, the inversion is immediate. Far from the source region Eq.\,\eqref{eq:GT1DFarfield} applies and
\begin{align}
E^{\rm tot}(z) 
	&= E^{\rm ext}(z)\! + \sum_{n,m} {G}(z\!-\!z_n) \mathcal{A}_{nm} E^{\rm ext}(z_m)
	\label{eq:totalfield}
	\\
	&= E_0 e^{i k z}\! + \sum_{n} \frac{\imath \beta \frac \Gamma 2 e^{\imath k |z-z_n|}}{\Delta - \imath [1 + \beta (N-1)] \Gamma/2 }.
\end{align}
In the forward direction ($z \gg z_N>0, z_n = 2 \pi k L$), we recover the transmitted power fraction \citep{Chang2012}
\begin{align}
\mathcal{T} =  \frac{|E^{\rm tot}|^2}{|E_0|^2}	
  =\left|1 + \frac{ \imath N   \beta \Gamma/ 2}{\Delta - \imath  [1 + \beta (N-1)]  \Gamma/ 2}\right|^2 .
  \label{eq:dbr_transmission}
\end{align}
At the single-body resonance $\Delta = 0$ we find
$%\begin{align}
\mathcal{T} = |1-r|^2$~, $\mathcal{R} = |r|^2$, $\mathcal{L} = 1 - \mathcal{R} - \mathcal{T}$, 
 $r = {N \beta}/[{N \beta + (1-\beta)}]~.
$ %\end{align}
We solve Eq.\,\eqref{eq:totalfield} numerically away from the Bragg resonance condition.

As long as Eq.\,\eqref{eq:farfieldmatrix} applies, a universal limit to the polariton spectrum  follows from the matrix norm (operator norm)
$%\begin{equation}
 ||\mathcal{A}^{-1}|| = \max\{|\epsilon - \imath \gamma/2 |\}) \equiv \max_n\left\{\sum_{m=1}^{N}|\mathcal{A}^{-1}_{nm}|\right\}.~
$%\end{equation}
The off-diagonal entries have an amplitude $\beta$, so that the norm suggests an extreme eigenvalue \mbox{$||\mathcal{A}^{-1}(\Delta = 0)|| =    [(N-1) \beta + 1] \Gamma/2= \gamma_+/2$}, exactly realized by the bright state at a Bragg resonance and confirming its universal character.

\subsection{Two-body resonances}
The dark and bright resonances of two-coupled emitters are obtained by direct diagonalization of the matrix $\mathcal{A}^{-1}$ for $N=2$, yielding
\begin{align}
\delta^{(2)}_\pm - \imath \gamma^{(2)}_{\pm}/2 = \Delta  - [\imath \Gamma/2 \pm G(\Delta z, \omega)]~.
\end{align}
Varying the emitter distance $0<\Delta z<\infty$, these eigenvalues cover the green curve in the complex plane in Fig.\,\ref{fig:specTBS}c. Similar states have been encountered in 3D ensembles \citep{Rusek2000, Skipetrov2014} but have more drastic impacts on linear arrangements. When two-body states dominate, $\mathcal{A}^{-1}$ approximately factorizes into a product of $2\times2$-matrices, which may be diagonalized individually so that each pair-submatrix gives rise to one pair of dark and bright states. The eigenvectors are $\Psi^{(2)}_\pm = (\pm 1, 1)/\sqrt{2}$.
At near-field coupling, the waveguide mode $E^{\rm ext}(z_n) \approx E_0 $ can only drive the symmetric eigenstate $\Psi_+$, inducing dipoles
\begin{equation}
d_n = \sum_{m=1}^2 \mathcal{A}_{nm} E_m = \frac{2 E_0}{\Delta - \imath \Gamma/2 - G(\Delta z)}.
\end{equation}
We find the pair-transmission amplitude
\begin{align}
T_2 %&= \lim_{z\to\infty}\left|E^{\rm ext}(z) / E_0(z) + \sum_{n=1}^N G(z - z_n) d_n\right|^2 \\
&= \left| 1 + \frac{2 \imath \beta \Gamma/2}{ \Delta - \imath \Gamma/2 - G(\Delta z)}\right|^2 = |1-r|^2.
\label{eq:two-body_extinction}
\end{align}
On the bright resonance ($\Delta - \Re[G(\Delta z)] \approx 0 $) and in near-field coupling regime ($\Im[G(\Delta z)] \approx \Gamma/2 $), $T_{2} = (1 - \beta)^2$ becomes identical to the one of a single resonant emitter.
Besides, the form of the two-body potential induces a correlation between line shift and linewidth: bright states are blue-shifted ($-\Delta =\delta>0$) with respect to the single resonance, and only these can provide significant extinction. From Eq.\,\eqref{eq:two-body_extinction} it is clear that the scattering is reduced and $r \to 0$ for red-shifted detuning $(-\Delta<0)$.

\subsection{Mean-field approximation}
In long dense chains ($kL\ll1\ll NkL$), we can rewrite Eq.\,(1) of the main text for the fields, assuming a homogeneous dipole line density $\rho(z)= N/Z$ over the system length $Z=NL$
\begin{equation}
E(z) = E_0(z) +\int_0^{Z}  dz' \alpha \rho(z') G(z-z') E(z')
\end{equation}
following the approach of Ref.\,\cite{Parola2014}. 
Since most of the emitters are coupled via the far field, we neglect near fields and use Eq.\,\eqref{eq:GT1DFarfield}. The good agreement with the numerical results justifies a posteriori the assumption that closeby emitters mainly induce a global frequency shift. Replacing the total field by a mean-field $E^{\rm tot}(z) = E_+ e^{i q z} + E_- e^{- i q z}$, the integrals can be performed, considering the discontinuity of $G$ at $z=z'$
\begin{widetext}
\begin{align}
E_+ e^{i q z} + E_- e^{- i q z} 
= E_0 e^{ikz} + \frac{N \beta}{Z} \alpha \Im[G(0)] %\times\\\nonumber&
\biggl[&
 -e^{iqz}\left(\frac{E_+ }{k-q}+ \frac{E_+ }{q+k}\right)
 -e^{-iqz}\left(\frac{E_- }{k-q}+\frac{E_- }{k+q}\right)\\\nonumber&
 + e^{-ikz}\left(\frac{E_+ e^{i (k+q)Z}}{k+q}+ \frac{E_- e^{i (k-q)Z}}{k-q}\right)%\\\nonumber&
 +e^{ikz}\left(\frac{E_- }{k+q}+\frac{E_+}{k-q}\right)
\biggr]
\end{align}
\end{widetext}
For self-consistency, the coefficients for each exponential must balance. Importantly, for $e^{\pm i q z}$ we have  
\begin{align}
1+\frac{N \beta}{Z} \alpha \Im[G(0)] \left(\frac{1}{k-q} + \frac{1}{k+q}\right) = 0~,
\end{align}
from which we obtain the effective wavenumber in the filled waveguide 
\begin{align}
\label{eq:disp}
q^2 = k^2 + 2 k \frac{N \beta}{Z} \alpha \Im[G(0)].
\end{align}
%The other equations allow to determine the values of the field amplitudes $E_+, E_-$.
We define a ``dressed'' effective mode index $\tilde n=q c/\omega = n_{\rm eff} q/k$ that fulfills a generalized Clausius-Mossotti relation \cite{Parola2014} when expanded at lowest order in $\alpha$
\begin{equation}
\frac{\tilde n -1}{\tilde n + 2} \approx \frac{2 \alpha}{3 k} \frac{N}{Z}  \beta\Im[G(0)]~.
\end{equation}
 To obtain the form given in Eq.\,(4) in the main article, we use 
$\beta \Im[G(0)] = \Gamma_0/\Gamma_{\rm bulk} \Im[G_{\rm bulk}(0)] \approx \frac{\omega/c}{2 \varepsilon_0  A_{\rm eff}}$ with the effective area 
$A_{\rm eff} = \frac{v_g n^2}{c}\frac{\int d^2 \vec{r} \varepsilon(\vec{r}) |\vec{E}_{\rm mode}(\vec{r})|^2}{\varepsilon(\vec{0})  |\vec{E}_{\rm mode}(\vec{0})|^2}$, the transverse field distribution of the mode  $E_{\rm mode}(\vec{r})$, and its group velocity  $v_g$ \citep{Jun2009}.
This underlines that the density needs to be averaged over the mode volume rather than the geometric fiber core.
Note that at this order, terms related to a local exclusion volume \citep{Parola2014} can be neglected. 

\vspace*{\fill}
\end{document}